\documentclass[useAMS,usenatbib]{mn2e}

\usepackage{tabularx}
\usepackage{xspace}
\usepackage{graphicx}
\newcommand{\ignore}[1]{}
\providecommand{\ao}{}

\renewcommand{\ao}{adaptive optics (AO)\renewcommand{\ao}{AO\xspace}\renewcommand{\Ao}{AO\xspace}\xspace}
\newcommand{\Ao}{Adaptive optics (AO)\renewcommand{\ao}{AO\xspace}\renewcommand{\Ao}{AO\xspace}\xspace}
\newcommand{\wfs}{wavefront sensor (WFS)\renewcommand{\wfs}{WFS\xspace}\renewcommand{\wfss}{WFSs\xspace}\xspace}
\newcommand{\wfss}{wavefront sensors (WFSs)\renewcommand{\wfs}{WFS\xspace}\renewcommand{\wfss}{WFSs\xspace}\xspace}
\newcommand{\shwfs}{Shack-Hartmann \wfs (SHWFS)\renewcommand{\shwfs}{SHWFS\xspace}\xspace}
\newcommand{\dm}{deformable mirror (DM)\renewcommand{\dm}{DM\xspace}\renewcommand{\dms}{DMs\xspace}\renewcommand{\Dms}{DMs\xspace}\renewcommand{\Dm}{DM\xspace}\xspace}
\newcommand{\dms}{deformable mirrors (DMs)\renewcommand{\dm}{DM\xspace}\renewcommand{\dms}{DMs\xspace}\renewcommand{\Dms}{DMs\xspace}\renewcommand{\Dm}{DM\xspace}\xspace}
\newcommand{\Dms}{Deformable mirrors (DMs)\renewcommand{\dm}{DM\xspace}\renewcommand{\dms}{DMs\xspace}\renewcommand{\Dms}{DMs\xspace}\renewcommand{\Dm}{DM\xspace}\xspace}
\newcommand{\Dm}{Deformable mirror (DM)\renewcommand{\dm}{DM\xspace}\renewcommand{\dms}{DMs\xspace}\renewcommand{\Dms}{DMs\xspace}\renewcommand{\Dm}{DM\xspace}\xspace}

\newcommand{\lqg}{linear-quadratic-gaussian (LQG)\renewcommand{\lqg}{LQG\xspace}\xspace}
\newcommand{\shs}{Shack-Hartmann sensor (SHS)\renewcommand{\shs}{SHS\xspace}\renewcommand{\shss}{SHSs\xspace}\xspace}
\newcommand{\shss}{Shack-Hartmann sensors (SHSs)\renewcommand{\shs}{SHS\xspace}\renewcommand{\shss}{SHSs\xspace}\xspace}
\newcommand{\lgs}{laser guide star (LGS)\renewcommand{\lgs}{LGS\xspace}\renewcommand{\lgss}{LGSs\xspace}\xspace}
\newcommand{\lgss}{laser guide stars (LGSs)\renewcommand{\lgs}{LGS\xspace}\renewcommand{\lgss}{LGSs\xspace}\xspace}
\newcommand{\Ngs}{Natural guide star (NGS)\renewcommand{\ngs}{NGS\xspace}\renewcommand{\Ngs}{NGS\xspace}\renewcommand{\ngss}{NGSs\xspace}\xspace}
\newcommand{\ngs}{natural guide star (NGS)\renewcommand{\ngs}{NGS\xspace}\renewcommand{\Ngs}{NGS\xspace}\renewcommand{\ngss}{NGSs\xspace}\xspace}
\newcommand{\ngss}{natural guide stars (NGSs)\renewcommand{\ngs}{NGS\xspace}\renewcommand{\Ngs}{NGS\xspace}\renewcommand{\ngss}{NGSs\xspace}\xspace}
\newcommand{\mems}{Micro-Electro-Mechanical Systems (MEMS)\renewcommand{\mems}{MEMS\xspace}\xspace}
\newcommand{\snr}{signal to noise ratio (SNR)\renewcommand{\snr}{SNR\xspace}\xspace}
\newcommand{\Moao}{Multi-object \ao (MOAO)\renewcommand{\moao}{MOAO\xspace}\renewcommand{\Moao}{MOAO\xspace}\xspace}
\newcommand{\moao}{multi-object \ao (MOAO)\renewcommand{\moao}{MOAO\xspace}\renewcommand{\Moao}{MOAO\xspace}\xspace}
\newcommand{\mcao}{multi-conjugate adaptive optics (MCAO)\renewcommand{\mcao}{MCAO\xspace}\xspace}
\newcommand{\ltao}{laser tomographic \ao (LTAO)\renewcommand{\ltao}{LTAO\xspace}\xspace}
\newcommand{\cpu}{central processing unit (CPU)\renewcommand{\cpu}{CPU\xspace}\renewcommand{\cpus}{CPUs\xspace}\xspace}
\newcommand{\cpus}{central processing units (CPUs)\renewcommand{\cpu}{CPU\xspace}\renewcommand{\cpus}{CPUs\xspace}\xspace}
\newcommand{\psf}{point spread function (PSF)\renewcommand{\psf}{PSF\xspace}\renewcommand{\psfs}{PSFs\xspace}\renewcommand{\Psf}{PSF\xspace}\xspace}
\newcommand{\psfs}{point spread functions (PSFs)\renewcommand{\psf}{PSF\xspace}\renewcommand{\psfs}{PSFs\xspace}\renewcommand{\Psf}{PSF\xspace}\xspace}
\newcommand{\Psf}{Point spread function (PSF)\renewcommand{\psf}{PSF\xspace}\renewcommand{\psfs}{PSFs\xspace}\renewcommand{\Psf}{PSF\xspace}\xspace}
\newcommand{\fpga}{field programmable gate array (FPGA)\renewcommand{\fpga}{FPGA\xspace}\renewcommand{\fpgas}{FPGAs\xspace}\xspace}
\newcommand{\fpgas}{field programmable gate arrays (FPGAs)\renewcommand{\fpga}{FPGA\xspace}\renewcommand{\fpgas}{FPGAs\xspace}\xspace}
\newcommand{\sor}{successive over-relaxation (SOR)\renewcommand{\sor}{SOR\xspace}\xspace}
\newcommand{\fdpcg}{Fourier domain pre-conditioned gradient (FDPCG)\renewcommand{\fdpcg}{FDPCG\xspace}\xspace}
\newcommand{\map}{maximum a-posteriori (MAP)\renewcommand{\map}{MAP\xspace}\xspace}

\newcommand{\elt}{Extremely Large Telescope (ELT)\renewcommand{\elt}{ELT\xspace}\renewcommand{\elts}{ELTs\xspace}\renewcommand{\eelt}{European ELT (E-ELT)\renewcommand{\eelt}{E-ELT\xspace}\xspace}\xspace}

\newcommand{\elts}{Extremely Large Telescopes (ELTs)\renewcommand{\elt}{ELT\xspace}\renewcommand{\elts}{ELTs\xspace}\renewcommand{\eelt}{European ELT (E-ELT)\renewcommand{\eelt}{E-ELT\xspace}\xspace}\xspace}

\newcommand{\eelt}{European Extremely Large Telescope (E-ELT)\renewcommand{\eelt}{E-ELT\xspace}\renewcommand{\elt}{ELT\xspace}\renewcommand{\elts}{ELTs\xspace}\xspace}

\newcommand{\dugall}{Durham University generalised adaptive optics laser laboratory (DUGALL)\renewcommand{\dugall}{DUGALL\xspace}\xspace}
\newcommand{\fwhm}{full-width at half-maximum (FWHM)\renewcommand{\fwhm}{FWHM\xspace}\xspace}
\newcommand{\wht}{William Herschel Telescope (WHT)\renewcommand{\wht}{WHT\xspace}\xspace}
\newcommand{\emccd}{electron multiplying CCD (EMCCD)\renewcommand{\emccd}{EMCCD\xspace}\renewcommand{\emccds}{EMCCDs\xspace}\xspace}
\newcommand{\emccds}{electron multiplying CCDs (EMCCDs)\renewcommand{\emccd}{EMCCD\xspace}\renewcommand{\emccds}{EMCCDs\xspace}\xspace}
\newcommand{\dasp}{Durham \ao simulation platform (DASP)\renewcommand{\dasp}{DASP\xspace}\renewcommand{\thedasp}{DASP\xspace}\renewcommand{\Thedasp}{DASP\xspace}\xspace}
\newcommand{\thedasp}{the Durham \ao simulation platform (DASP)\renewcommand{\dasp}{DASP\xspace}\renewcommand{\thedasp}{DASP\xspace}\renewcommand{\Thedasp}{DASP\xspace}\xspace}
\newcommand{\Thedasp}{The Durham \ao simulation platform (DASP)\renewcommand{\dasp}{DASP\xspace}\renewcommand{\thedasp}{DASP\xspace}\renewcommand{\Thedasp}{DASP\xspace}\xspace}
\newcommand{\mpi}{Message Passing Interface (MPI)\renewcommand{\mpi}{MPI\xspace}\xspace}
\newcommand{\smp}{symmetric multi-processing (SMP)\renewcommand{\smp}{SMP\xspace}\xspace}
\newcommand{\svd}{singular value decomposition (SVD)\renewcommand{\svd}{SVD\xspace}\xspace}
\newcommand{\gpu}{graphics processing unit (GPU)\renewcommand{\gpu}{GPU\xspace}\renewcommand{\gpus}{GPUs\xspace}\xspace}
\newcommand{\gpus}{graphics processing units (GPUs)\renewcommand{\gpu}{GPU\xspace}\renewcommand{\gpus}{GPUs\xspace}\xspace}
\newcommand{\fft}{fast Fourier transform (FFT)\renewcommand{\fft}{FFT\xspace}\xspace}
\newcommand{\ifu}{integral field unit (IFU)\renewcommand{\ifu}{IFU\xspace}\xspace}
\newcommand{\darc}{the Durham \ao real-time controller (DARC)\renewcommand{\darc}{DARC\xspace}\renewcommand{\Darc}{DARC\xspace}\xspace}
\newcommand{\Darc}{The Durham \ao real-time controller (DARC)\renewcommand{\darc}{DARC\xspace}\renewcommand{\Darc}{DARC\xspace}\xspace}
\newcommand{\cots}{commercial off-the-shelf (COTS)\renewcommand{\cots}{COTS\xspace}\xspace}
\newcommand{\rtcp}{real-time control pipeline (RTCP)\renewcommand{\rtcp}{RTCP\xspace}\xspace}
\newcommand{\rms}{root-mean-square (RMS)\renewcommand{\rms}{RMS\xspace}\xspace}
\newcommand{\sFPDP}{serial Front Panel Data Port (sFPDP)\renewcommand{\sFPDP}{sFPDP\xspace}\xspace}
\newcommand{\wpu}{wavefront processing unit (WPU)\renewcommand{\wpu}{WPU\xspace}\xspace}

\newcommand{\rtcs}{real-time control system (RTCS)\renewcommand{\rtcs}{RTCS\xspace}\renewcommand{\rtcss}{RTCSs\xspace}\xspace}
\newcommand{\rtcss}{real-time control systems (RTCSs)\renewcommand{\rtcs}{RTCS\xspace}\renewcommand{\rtcss}{RTCSs\xspace}\xspace}
\newcommand{\eso}{European Southern Observatory (ESO)\renewcommand{\eso}{ESO\xspace}\renewcommand{\theeso}{ESO\xspace}\xspace}
\newcommand{\theeso}{\renewcommand{\theeso}{ESO\xspace}the \eso}
\newcommand{\scao}{single conjugate \ao (SCAO)\renewcommand{\scao}{SCAO\xspace}\renewcommand{\Scao}{SCAO\xspace}\xspace}
\newcommand{\Scao}{Single conjugate \ao (SCAO)\renewcommand{\scao}{SCAO\xspace}\renewcommand{\Scao}{SCAO\xspace}\xspace}
\newcommand{\glao}{ground layer \ao (GLAO)\renewcommand{\glao}{GLAO\xspace}\xspace}
\newcommand{\eagle}{ELT Adaptive optics for GaLaxy Evolution (EAGLE)\renewcommand{\eagle}{EAGLE\xspace}\xspace}
\newcommand{\maory}{multi-conjugate \ao relay for the \eelt (MAORY)\renewcommand{\maory}{MAORY\xspace}\xspace}
\newcommand{\muse}{Multi Unit Spectroscopic Explorer (MUSE)\renewcommand{\muse}{MUSE\xspace}\xspace}
\newcommand{\vlt}{Very Large Telescope (VLT)\renewcommand{\vlt}{VLT\xspace}\xspace}

\newcommand{\tmt}{Thirty Metre Telescope (TMT)\renewcommand{\tmt}{TMT\xspace}\xspace}
\newcommand{\xao}{eXtreme \ao (XAO)\renewcommand{\xao}{XAO\xspace}\xspace}

\newcommand{\vla}{Very Large Array (VLA)\renewcommand{\vla}{VLA\xspace}\xspace}
\newcommand{\jwst}{{\em James Webb Space Telescope} \citep[JWST,][]{jwst}\renewcommand{\jwst}{{\em JWST}\xspace}\xspace}
\newcommand{\hst}{{\em Hubble Space Telescope (HST)}\renewcommand{\hst}{{\em HST}\xspace}\xspace}
\newcommand{\ifss}{integral-field spectrographs (IFSs)\renewcommand{\ifss}{IFSs\xspace}\renewcommand{\ifs}{IFS\xspace}\xspace}
\newcommand{\ifs}{integral-field spectrograph (IFS)\renewcommand{\ifss}{IFSs\xspace}\renewcommand{\ifs}{IFS\xspace}\xspace}
\newcommand{\ifus}{integral field units (IFUs)\renewcommand{\ifus}{IFUs\xspace}\xspace}
\newcommand{\mos}{multi-object spectrograph (MOS)\renewcommand{\mos}{MOS\xspace}\xspace}
\newcommand{\goodss}{Great Observatories Origins Deep Survey (GOODS)-S\renewcommand{\goodss}{GOODS-S\xspace}\xspace}
\newcommand{\goods}{Great Observatories Origins Deep Survey (GOODS)\renewcommand{\goods}{GOODS\xspace}\xspace}
\newcommand{\scmos}{scientific CMOS (sCMOS)\renewcommand{\scmos}{sCMOS\xspace}\xspace}
\newcommand{\aof}{Adaptive Optics Facility (AOF)\renewcommand{\aof}{AOF\xspace}\xspace}

\usepackage{amssymb}

\title[ELT MOAO performance in deep fields]{Wide-field adaptive optics
  performance in cosmological deep fields for multi-object
  spectroscopy with the European Extremely Large Telescope}

\author[A.\ G.\ Basden et al.]{A.\ G.\ Basden,$^{1}$\thanks{E-mail:
    a.g.basden@durham.ac.uk (AGB)} C.\ J.\ Evans$^2$ and T.\ J.\ Morris$^1$\\
$^{1}$Department of Physics, South Road, Durham, DH1 3LE, UK\\
$^2$UKATC, Royal Observatory, Blackford Hill, Edinburgh, EH9 3HJ, UK}

\begin{document}
\maketitle

\begin{abstract}
A multi-object spectrograph on the forthcoming European Extremely
Large Telescope will be required to operate with good sky coverage.
Many of the interesting deep cosmological fields were deliberately
chosen to be free of bright foreground stars, and therefore are
potentially challenging for adaptive optics (AO) systems.  Here we
investigate multi-object AO performance using sub-fields chosen at
random from within the Great Observatories Origins Deep Survey
(GOODS)-S field, which is the worst case scenario for five deep
fields used extensively in studies of high-redshift galaxies.  Our AO
system model is based on that of the proposed MOSAIC instrument but
our findings are equally applicable to plans for multi-object
spectroscopy on any of the planned Extremely Large Telescopes.
Potential guide stars within these sub-fields are identified and used
for simulations of AO correction.  We achieve ensquared energies
within 75~mas of between 25-35\% depending on the sub-field, which is
sufficient to probe sub-kpc scales in high-redshift galaxies.  We also
investigate the effect of detector readout noise on AO system
performance, and consider cases where natural guide stars are used for
both high-order and tip-tilt-only AO correction.  We also consider how
performance scales with ensquared energy box size.  In summary, the
expected AO performance is sufficient for a MOSAIC-like instrument,
even within deep fields characterised by a lack of bright
foreground stars.
\end{abstract}


\section{Introduction}
\label{sect:intro}

One of the primary scientific motivations for the forthcoming \elts,
which will have primary mirror diameters in excess of 20~m, is to
understand the evolution of galaxies -- from formation of the most
distant systems known, seen only a few million years after the Big
Bang, through to disentangling the structural components and histories
of the nearby galaxies that we see today.

Many of the breakthroughs in galaxy studies over the past twenty years
have been enabled by observations of selected ``deep fields'' on the
sky. The core datasets for these efforts have been optical and
near-infrared imaging from long integrations with the \hst and
ground-based observatories, complemented by a wealth of
multi-wavelength information from facilities such as {\em Chandra},
{\em Spitzer}, {\em Herschel} and the \vla.

Given the substantial observational investment in these fields, and
the richness of the multi-wavelength data available, they will almost
certainly be the target of future programmes with both the \jwst and
the \eelt \citep{eelt}.  Specifically, many of the sources known in
these fields are sufficiently faint that they are beyond our current
spectroscopic capabilities, and follow-up spectroscopy will require the
improved sensitivities of these exciting new facilities.

One of the most important new techniques for studies of high-redshift
galaxies over the past decade has been the development of \ifss on
8-10\,m class telescopes.  These have enabled {\em spatially-resolved}
studies of galaxies out to a redshift of $z$\,\,$\sim$\,3 \citep[see,
  e.g., the review by][]{Glazebrook2013}. In particular, the use of
\ao with \ifs instruments has provided unprecedented views of the
substructure and physical properties of high-$z$ galaxies \citep[see,
  e.g., the comparisons of AO-corrected and seeing-limited
  observations from][]{Newman2013short}.

However, at $z$\,$\gtrsim$\,1.5 we are currently limited to
spatially-resolved spectroscopy of only the most luminous and/or
massive galaxies; to observe a representative sample of the galaxy
population in the early Universe we require the spectroscopic
sensitivity of the \eelt.  Moreover, to compile large samples (of
thousands) of objects within a realistic observing time, we require the
combination of multiple \ifus and a wide-field \ao system.

The technical requirements for such observations, in particular the
necessary AO performances, were presented by \citet{Puech2008,
  Puech2010}. In brief, to probe scales of $\sim$1\,kpc at high
redshift \citep[e.g.,][]{2007ApJ...670..237B} we require ensquared energies
of 20 to 30\% in $\sim$75\,mas in the $H$-band. Coarser sampling of
$\sim$100 to 150\,mas (with comparable ensquared energy) is sufficient
for the recovery of large-scale dynamics in the target galaxies. These
requirements strongly influenced the conceptual design for the
proposed EAGLE instrument \citep{Cuby2010short}, and they are now
shaping the design of the MOSAIC concept \citep{Evans2014, Hammer2014}
for a \mos for the \eelt. As noted above, many of the potential
targets for \mos observations with any of the planned \elts will
likely be located in the existing extragalactic deep fields.  To
achieve the \moao correction for an instrument such as MOSAIC we need
\ngss within the patrol field of the instrument, but a key feature of
the deep fields was that they were deliberately chosen to be free of
relatively bright foreground stars (to avoid problems relating to
saturation, diffraction, persistence, etc.).  For example, the lack of
suitable guide stars ($V<14$) within a $2'\times2'$ field for
observations with the Multi-Conjugate Adaptive Optics Demonstrator
 \citep[MAD,][]{marchetti2007}, severely limited its use to observe
cosmological deep fields.  

In the following we therefore investigate whether we can obtain
sufficient image quality for spatially-resolved spectroscopy of
high-$z$ galaxies in one of the most important extragalactic deep
fields, the \goodss field \citep{Giavalisco2004short}. The \goodss field is
one of five observed as part of the Cosmic Assembly Near-infrared
Deep Extragalactic Legacy Survey \citep[CANDELS;][]{Grogin2011short},
providing unprecedented depth and wavelength coverage for such a large
area of \hst imaging.  In addition to the rich photometric
catalogues, there has been significant spectroscopic follow-up of
galaxies in the \goodss field \citep[e.g.][]{Popesso2009}, and this
will be extended by the recently approved VANDELS ESO Public
Spectroscopic Survey (PIs: R.~McLure \& L.~Pentericci).  Thus, in the
coming years we will have considerable information in the \goodss
region to select well-defined samples of target galaxies for
spatially-resolved spectroscopy with the \elts.

In addition to considering the \goodss field, we have performed a
study of other fields including GOODS-N, UDS, EGS and COSMOS
\citep{Grogin2011short}.  Within each of these fields, we have taken a
random sample of ten 10~arcminute sub-fields and determined the number
of stars within these fields with a $r'$ magnitude of brighter than 16,
and thus available for high order wavefront sensing (i.e.\ delivering
more than a few photons per sub-aperture per frame).
Table~\ref{tab:starcount} shows the number of suitable \ngss within
these sub-fields; the \goodss field provides the fewest guide stars
within its sub-fields.  We have therefore chosen these sub-fields for
further study as the pessimistic case: our results are equally
applicable for the other fields.

\begin{table}
\centering
\begin{minipage}{7cm}
\caption{NGS availability within ten random 10~arcminute sub-fields
  from within the given cosmological deep fields. The numbers
  presented are the number of sub-fields containing only $N$ natural
  guide stars with $r'<16$ mag.}
\label{tab:starcount}
\begin{tabularx}{\linewidth}{llllllllll}
Field &\multicolumn{9}{c}{Sub-fields containing only $N$ guide stars}\\
\multicolumn{2}{r}{$N$= 2} &3&4&5&6&7&8&9&$>$9\\ \hline
GOODS-S&1&2&2&4&1&0&0&0&0\\
GOODS-N&0&1&2&1&2&2&1&1&0\\
UDS    &0&0&0&0&0&2&3&3&2\\
EGS    &0&1&0&0&2&1&2&1&3\\
COSMOS &0&0&0&0&0&0&0&0&10\\
\end{tabularx}
\end{minipage}
\end{table}

The NFIRAOS system on the \tmt has a science goal of 50\% sky
coverage, and is expected to deliver \ao correction over most of the
observable sky for 50\% of fields \citep{2011aoel.confP..18A}.  The
increased light collecting area of the \eelt will further improve
field observability.  Since the \goodss field is looking directly out
of the Milky Way galaxy, it is a pessimistic case.  Indeed, a
statistical approach to sky coverage for the \eelt by \eso
\citep{calamida2009} found that, in a typical field at the latitude of
the \goodss field, we would expect more guide stars than we find in
this field (it is pessimistic), and that even at the
Galactic pole we would expect at least a couple of stars.  Therefore
we are confident that the fields under consideration here are relevant
for the whole sky.

In \S2, we introduce the \ngs fields that we investigate, and provide
details of the \ao modelling that we perform.  In \S3 we discuss our
results, and we conclude in \S4.

\section{Modelling of ELT Adaptive Optics in the GOODS-S field}
Within the \goodss Deep\,$+$\,Wide\,$+$\,ERS survey region from
CANDELS we generated ten random 10$'$ diameter fields to investigate
potential \ngs asterisms.  Using the data assembled for the fourth
United States Naval Observatory CCD Astrograph Catalog
\citep[UCAC4,][]{Zacharias2013}, we recovered the positions and
  magnitudes of stars within each field (which are complete down to
  $R$\,$\sim$\,16\,mag), and used these as inputs to investigate the
  range of simulated \ao performances.  Specifically, we use the
  cross-matched Sloan $r'$-band magnitudes (in the AB system) from the
  American Association of Variable Star Observers (AAVSO) Photometric
  All-Sky Survey (APASS).

Fig.~\ref{fig:asterism} shows the fields that we have investigated,
giving the \ngss available within these fields and their corresponding
$r'$-band magnitudes.  For reference, we also show the notional \lgs
positions.  The field centres are given in
Table~\ref{tab:field}, and the overlap of the fields is shown graphically in
Fig.~\ref{fig:overlap}.  A key point of this study is that of random
field selection, and by estimating \ao performance in randomly
selected sub-fields, we are able to demonstrate the availability of
suitable \ngs targets over the whole of the \goodss field.  The
alternative approach would be to determine all the \ngs asterisms
providing good \ao performance within the \goodss field.  However, it
would then be necessary to accept that, if a scientific target lay
just outside the corrected field of view, it would not be observable,
or at least, the \ao correction would be lower than required.

The faintest target that we consider has an $r'$ magnitude of 16.3,
which we translate to about three detected photons per sub-aperture
per frame.  We do not consider the use of any fainter targets as too
few photons would be received.  In the cases where \ngss are used for
tip-tilt correction only, it would be possible to further reduce the
source flux, though we do not investigate this here since the
lack of high-order \ngs information then leads to reduced \ao
performance (even if the tip-tilt stars are very bright).

\begin{figure*}
\includegraphics[width=\linewidth]{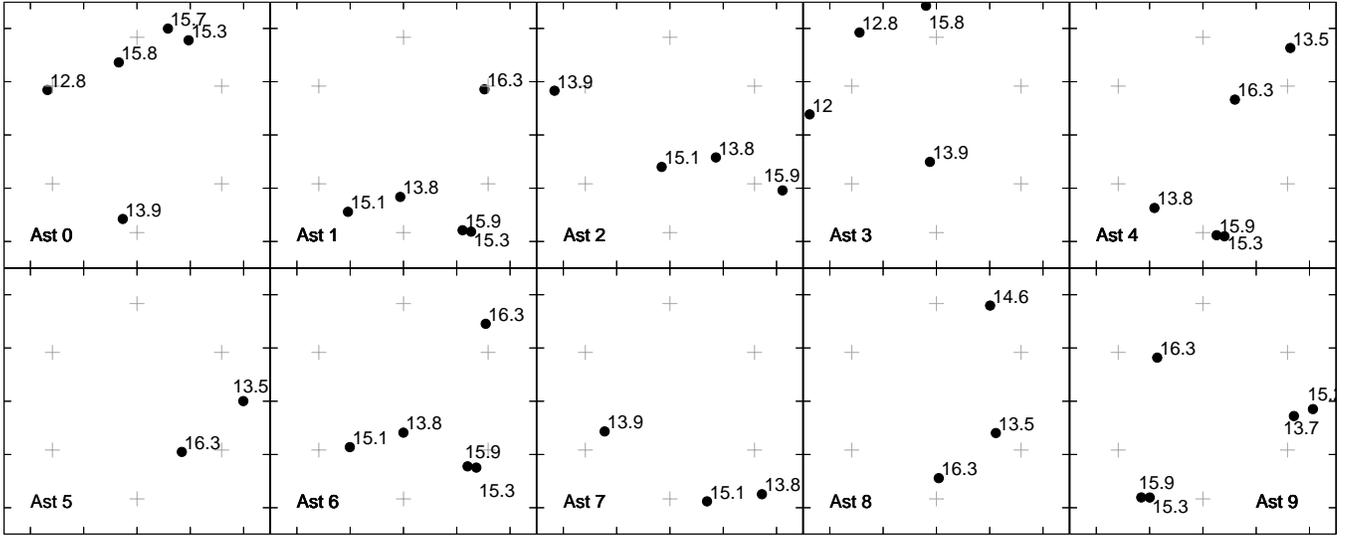}
\caption{The ten NGS asterisms used to investigate potential AO
  performance in the GOODS-S field.  Each sub-field is 10~arcminutes in diameter.
The numbers within these figures provide the corresponding guide star
$r'$-band magnitudes, and the grey crosses show LGS positions.}
\label{fig:asterism}
\end{figure*}

\begin{table}
\centering
\begin{minipage}{7cm}

\caption{Central coordinates for the 10$'$ fields in which we have
  investigated the AO performances (J2000).}
\label{tab:field}
\begin{tabularx}{\linewidth}{ccc}
Asterism & RA / degrees & Dec / degrees\\ \hline
0 & 53.10850525 & -27.71668243\\
1 & 53.16623306 & -27.81152534\\
2 & 53.19574356 & -27.77984810 \\
3 & 53.14884949 & -27.71123886\\
4 & 53.16050720 & -27.84217262\\
5 & 53.10297775 & -27.83361053\\
6 & 53.18798828 & -27.80929756\\
7 & 53.14844513 & -27.74827385\\
8 & 53.08246613 & -27.86190033\\
9 & 53.16571426 & -27.89375114\\
\end{tabularx}
\end{minipage}
\end{table}
\begin{figure}
\includegraphics[width=\linewidth]{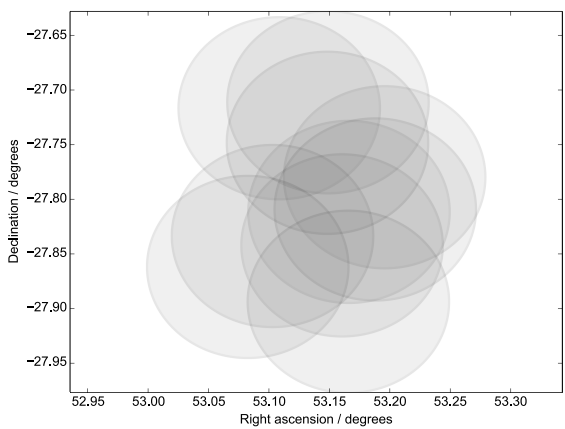}
\caption{Positions of the ten random fields used to investigate
potential NGS asterisms, within the GOODS-S field.  The circles
represent the 10 arcminute field-of-view of the telescope at each location.}
\label{fig:overlap}
\end{figure}

\subsection{Adaptive optics simulations}
We use a Monte-Carlo simulation tool \citep{basden5}, \thedasp, to
provide performance estimates for a \moao-corrected \mos on the \eelt,
using the \ngs asterisms defined in Fig.~\ref{fig:asterism}.  The
model has the same basic form as that used for performance modelling
of the EAGLE concept \citep{basden8}, and previous analysis of \eelt
\moao performances \citep{basden12,basden15}.  Although the exact
details of the MOSAIC concept are still under study (e.g.\ number and
final specification of science channels), our objective here was to
investigate the broader aspects of a \moao-corrected \mos, so we
investigate a wide range of \ngs magnitudes (\S\ref{sect:scaling}),
detector performances (\S\ref{sect:rn}), tip-tilt correction
(\S\ref{sect:tt}), and spatial element size on-the-sky
(\S\ref{sect:ee}).  We note that the \moao performance across the
telescope field-of-view has been investigated previously
\citep{basden12}, so here we provide on-axis performance estimates (at
the centre of the \lgs asterism) to simplify our results.

In summary, we use six sodium \lgss (589~nm wavelength) equally spaced
around the edge of a 7.3~arcminute diameter circle (the widest \lgs
asterism that can be transported to the \eelt focal plane), each with
$74\times74$ sub-apertures and $16\times16$ pixels per sub-aperture.
The telescope diameter is taken to be 38.5~m, and has a pupil function
taken to match that of the multi-hexagon \eelt.  The secondary mirror
obscuration is about 11~m in diameter, and the \dm pitch is 52~cm.  We
have used a set of 35~layer atmospheric profiles, available on request
from \theeso.  This profile is the result of years of data collection
at the Paranal Observatory, and has an outer scale of $L_0=25$~m, and
a Fried's parameter of $r_0=15.7$~cm defined at zenith.  In the
simulations presented here, we assume that observations are made at 30
degrees from zenith.  We include the cone effect (due to the finite
distance to the \lgs spots) and spot elongation, assuming a
sodium-layer profile with a Gaussian shape centred at 90~km with a
5~km \fwhm.  We use up to five \ngss, depending upon availability, and
these are also sampled by $74\times74$ Shack-Hartmann wavefront sensor
sub-apertures, which operate at the centre of the $r'$-band (625~nm).
We do not take \ngs chromatic effects into consideration.  We include
photon shot noise in our \wfss, and our default case includes no
readout noise, though we investigate the effect on performance that
this has in \S\ref{sect:rn}.  In recent years, both \emccd and \scmos
technologies have progressed significantly, and so it is likely that
very low readout noise levels will be achievable within the time frame
of \elt instrumentation.  \emccds \citep{basden1} routinely achieve
0.1~photo-electrons readout noise (with further reductions possible
depending on their mode of operation), and \scmos technology now
achieves levels as low as 0.9~photo-electrons.  For simplicity, we use
a basic centre-of-gravity algorithm to estimate wavefront slope for
both \lgs and \ngs \wfss, even though it has previously been shown
that other algorithms can yield better performance, for example using
correlation methods \citep{basden14}.

The \eelt M4 deformable mirror is assumed to be conjugated to the
ground-layer turbulence, and has $75\times75$ actuators.  We
investigate the $H$-band on-axis performance and use a \moao \dm with
$75\times75$ actuators.  Unless otherwise stated, our primary performance
metric is the percentage of the \psf ensquared energy within 75~mas,
chosen to match the scales of interest in high-$z$ galaxies
\citep{Puech2008}.  A
tomographic reconstruction of the atmospheric turbulence is performed
using a minimum variance formulation, with phase covariance
approximated by a Laplacian function.  Noise covariance is assumed to
be constant for each \wfs, and dependent on guide star signal level.
Wavefront reconstruction involves a virtual \dm formulation, and we
use 12 virtual \dms, having found that there is insignificant
performance gain when additional \dms (with associated increased
simulation complexity) are introduced with the 35 layer atmosphere model.  The reconstructed phase is
then projected onto the physical \dms, which perform the \ao
correction.  We model \dms using an interpolated spline function,
which gives a good fit to surface models of most known \dm types.

We assume an \ao system update rate of 250~Hz, a baseline for MOSAIC,
since the \moao \dms will be operated in open-loop, and simulate 20~s of
telescope time, verifying that the \psfs have converged.  We note that
the CANARY \moao demonstration instrument \citep{canaryshort} can also
be operated at 250~Hz.

A previous study has shown that \ao correction for the \lgs
configuration considered here is relatively constant over the
field-of-view, differing by only a few percent in ensquared energy
\citep{basden12}.  Here, we therefore concentrate on the on-axis
direction which is furthest from the \lgs locations.

\subsubsection{Guide star signal level}
We base our \ngs signal levels on r$'$ magnitudes, centred at $\lambda=625$~nm
with a $\delta\lambda=140$~nm bandwidth \citep{1996AJ....111.1748F}.  At magnitude 0,
the r$'$ band gives a flux, $F$ of 3631~Jy \citep{1983ApJ...266..713O}, and
therefore a zero-magnitude star gives
$1.23\times10^{10}$~photons~m$^{-2}$~s$^{-1}$.  We assume a telescope
throughput of $T_{tel}=90$\% \citep[see e.g.,][]{Puech2010}, and a \wfs
throughput of $T_{wfs}=85$\%, giving a final flux estimation equal to:
\ignore{
\begin{equation}
s=3631\times1.51\times10^7\times\frac{140}{625}\times
10^{-0.4r'}\frac{0.5^2}{250}\times0.9\times 0.85
\label{eq:sig}
\end{equation}
}
\begin{equation}
s=F\times J\times
10^{-0.4r'}\frac{A}{f}\times T_{tel}\times T_{wfs}
\label{eq:sig}
\end{equation}

where $s$ is the \ngs signal in photons per sub-aperture per frame,
and $r'$ is the guide-star magnitude.  $J=1.51\times10^7$ photons per
second per square metre per fractional bandwidth ($\frac{\delta\lambda}{\lambda}$).
$A$ is the sub-aperture area, $0.5^2$~m$^2$, and $f$ is the \ao frame
rate of 250~Hz.  

Because there are many uncertainties in this flux calculation, we also
investigate simulation performance as a function of a scaling of this
signal level, allowing estimates to be updated once factors such as
actual guide star observation band (and bandwidth, since \wfss can be
very broadband with appropriate atmospheric dispersion correction) and
detector quantum efficiency are known.  

It should be noted that these signal levels are very faint.  The
faintest star in the \ngs asterisms that we investigate has a r$'$
magnitude of 16.3, giving about 3 photons per sub-aperture per frame,
according to Eq.~\ref{eq:sig}.  We take no special measures with such
faint guide stars: they are processed in the same way as all others.
However, because we use a maximum apriori wavefront reconstruction
algorithm, \wfs noise is taken into consideration, and
signals from noisier wavefront sensors are penalised.  Guide star
signal levels could be increased by reducing the \wfs frame rate.
However, we do not consider this option here, partly because we are
unable to easily operate \lgs and \ngs \wfss at different rates in our
simulations, and because the improvement in performance would only be
slight, for a large gain in complexity:  we already take \wfs noise
into consideration in our wavefront reconstruction, and because the
correction of the high spatial frequencies using high-order \wfss
requires high time resolution, with reducing effectiveness as frame
rate decreases.  However, it should be noted that a reduction in \ngs
frame rate is likely to lead to a small improvement in \ao performance
when the faintest \ngss are considered.

We assume that \lgs signal levels are not photon limited \citep{2010SPIE.7736E..28H}.

\section{MOAO performance in the GOODS-S field}
\label{sect:scaling}
Evaluating the \ao performance for each asterism at the reference
signal level alone will only give a performance snapshot for this
signal level.  We therefore consider increasing and reducing the
signal levels across the board (i.e.\ for all \ngs), for each asterism
under consideration.  Fig.~\ref{fig:sig} shows that the signal levels
available for guide stars within the \goodss field are indeed low, and
that an order of magnitude increase in flux (2.5 astronomical
magnitudes) is required to ensure that there is then little
performance gain by further increasing the signal levels.  However, it
also shows that at the reference flux levels, between 25-35\%
ensquared energy within 75~mas can be achieved (this increases to
30-38\% when flux is not the limiting factor).  The uncertainties
within the model (based on repeat simulations) for this, and the
following, figures are at the 1\% level (not shown in the figures to
aid clarity).

\begin{figure}
\includegraphics[width=\linewidth]{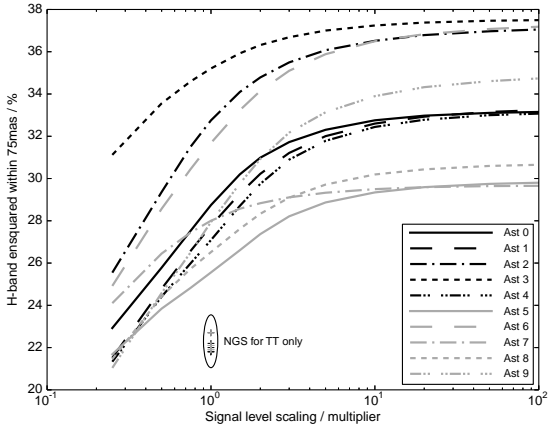}
\caption{AO performance (ensquared energy) as a function of NGS
  asterism signal level scaling (with all guide star signals scaled by the
  X-axis value).  For reference, the range of performance achieved
  when NGS information is used for tip-tilt correction only is also
  shown, and gives constant \ao performance between signal level
  scales from one to 100. The legend provides the asterism number (ast).}
\label{fig:sig}
\end{figure}

Fig.~\ref{fig:fwhm} shows the \ao corrected \psf \fwhm as a function
of signal level, compared with the theoretical diffraction limit for a
38.5~m aperture.  For the unmodified signal level (a scaling factor of
unity), all asterisms lead to a \psf with a \fwhm less than twice that
of the theoretical, with some being only 10\% larger.  The \psfs
themselves are shown in Fig.~\ref{fig:psf} for the case of default
signal level.  All the \psfs are well constrained, however, in the
case of asterism 5, significant structure is displayed due to the
presence of only two guide stars (one of which being very faint).  

\begin{figure}
\includegraphics[width=\linewidth]{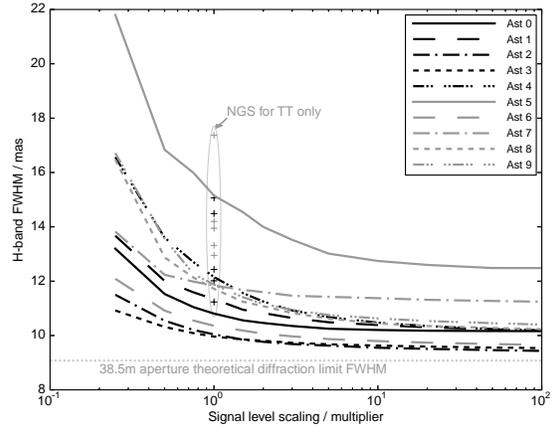}
\caption{AO performance (FWHM) as a function of NGS
  asterism signal level scale (with all guide star signals scaled by the
  X-axis value).  For reference, the range of performance achieved
  when NGS information is used for tip-tilt correction only is also
  shown, and gives constant \ao performance between signal level
  scales from one to 100.  The theoretical aperture limit is also
  shown, corresponding to the FWHM of a diffraction-limited spot on a
  38.5~m aperture.}
\label{fig:fwhm}
\end{figure}

\begin{figure*}
\includegraphics[width=\linewidth]{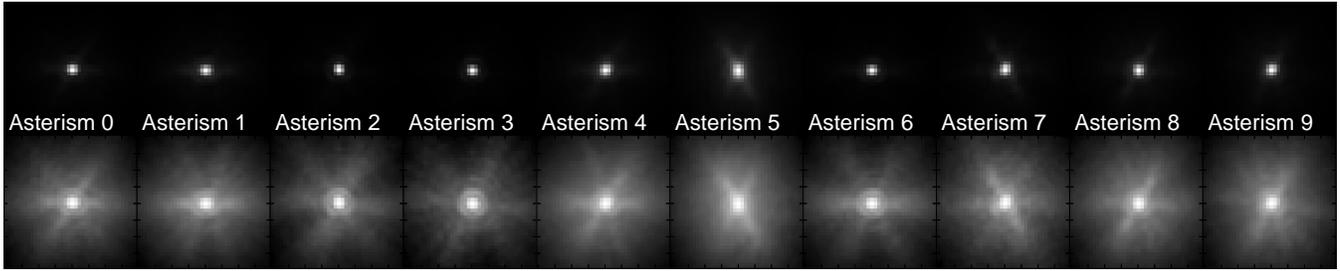}
\caption{Simulated AO corrected H-band PSFs for the ten asterisms under
  consideration here.
  Each PSF box has an edge size of 177~mas.  The top row shows the
  PSF, while the bottom row shows the natural logarithm of the PSF,
  showing the underlying hexagonal structure due to the LGS arrangement.
}
\label{fig:psf}
\end{figure*}

\subsection{The effect of readout noise on AO performance}
\label{sect:rn}
We have so far assumed that our detectors have no readout noise, with
photon shot noise being the only \wfs noise source.  We now consider
the \ao performance when \ngs readout noise is introduced.  We have
considered readout noise levels based on current detector
technologies, ranging from 0.01 photo-electrons to 1 photo-electron.   
Fig.~\ref{fig:noise} shows the \ao performance as a function of readout
noise for the different \ngs asterisms under consideration.
Here we can see that noise can have a significant effect on
performance, and therefore careful consideration should be given to
the detector technology (\emccd or \scmos) used.  

We have not considered detector quantum efficiency or the excess noise
factor introduced by \emccds (which can effectively halve the quantum
efficiency), though the effect that this has on performance can be
garnered from Fig.~\ref{fig:sig}: it can be seen that with a signal
level scaling of 0.5 (due to the \emccd excess noise factor), the \ao
performance is higher than that with a readout noise of 1
photo-electron in Fig.~\ref{fig:noise} (corresponding to an \scmos
detector).  This suggests that \emccd technology is more appropriate
here.

\begin{figure}
\includegraphics[width=\linewidth]{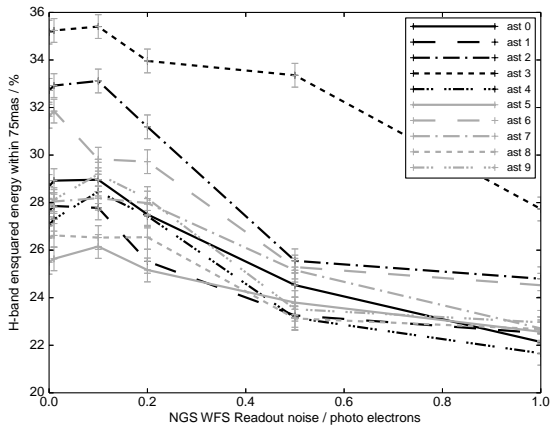}
\caption{AO performance as a function of NGS
  readout noise, for the different asterisms under consideration here.}
\label{fig:noise}
\end{figure}

Detector readout noise is key to the \ao performance.  However, our
treatment of the effect of noise has been basic: we have simply
removed a background level after adding readout noise (with a random
Gaussian distribution) and relied on the minimum variance wavefront
reconstruction.  We have ignored sky background, which is low
in the R$'$-band at 250~Hz frame rates.  We have not
considered more advanced techniques often employed in \ao, for example
pixel calibration based on brightest pixels within a sub-aperture
\citep{basden10}, or Gaussian noise removal algorithms.  The baseline
for the EAGLE concept \citep{2008SPIE.7014E..53Cshort}, a forerunner
for MOSAIC, was to use Shack-Hartmann wavefront sensors, and we have
therefore not considered other sensor types.  We have also not
considered the benefits that could be obtained by reducing the number
of \ngs sub-apertures.  Further, we have not considered the change in
performance if the faintest stars within an asterism are disregarded
(i.e.\ not used for \ao), which may reduce noise propagation and hence
increase \ao performance.  Therefore it is likely that performance
improvements could be realised when using noisy detectors over the
results presented here, which can therefore be taken as pessimistic.
The upper bound to performance, however, is given by the high flux
cases in Fig.~\ref{fig:sig}.

\subsection{NGS for tip-tilt correction only}
\label{sect:tt}
Given that the \ngs signal levels are very low for many of the guide stars
available within the selected \goodss sub-fields, it is worth
considering the \ao performance obtained when the \ngs are used for
tip-tilt correction only.  Even using the faintest guide stars
available provides ample signal for tip-tilt estimation, with
thousands of photons collected across the telescope aperture.
However, this then results in a lack of tomographic wavefront
information, since the wide \lgs asterism is unable to fully sample
the turbulent volume.  We find that the \ao performance is reduced to 
about 22-23\% ensquared energy within 75~mas, irrespective of the \ngs
asterism used (Fig.~\ref{fig:sig}).  \ngs tip-tilt-only correction
within the \goodss field is therefore not optimal: the high-order information obtainable from the \ngss is valuable.  We also
find that there is no increase in performance when guide star flux
is increased by up to a factor of 100 beyond the nominal flux
(levelling off at about 10 times flux, i.e.\ \ngss which are 2.5 magnitudes
brighter than those studied here).

\subsection{Ensquared energy diameter}
\label{sect:ee}
The fractional ensquared energy as a function of spatial size on the
sky is shown in Fig.~\ref{fig:enc}.  We have shown the best and worst
asterisms in terms of performance (asterisms 5 and 7, respectively), and
provide results for the default case, for an increase in the \ngs
signal by a factor of ten (where the \ao performance approaches that of
the high light-level case), and for when no high-order \ngs information
is used (i.e., the \ngs are only used to determine the required
tip-tilt correction). These results will directly inform design
decisions for the MOSAIC concept, and their trends will be relevant
to discussions of \ifu spaxel size verses expected performance for other
future instruments.

\begin{figure}
\includegraphics[width=\linewidth]{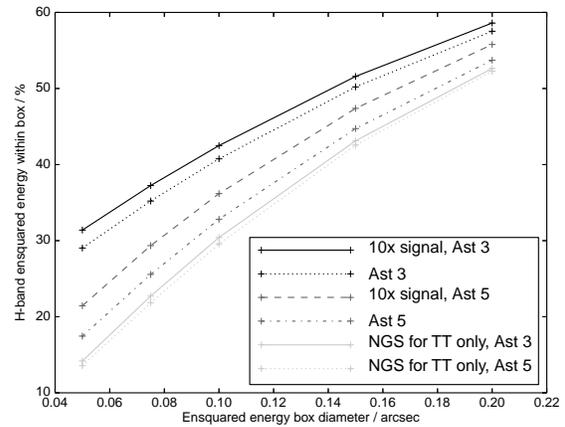}
\caption{The fraction of PSF energy ensquared as a
  function of box size.  As shown in the legend, results are for the
  best (3) and worse (5) performing NGS asterisms, for three separate
  cases:  when the NGS signal level is increased by a factor of ten, for
  the default NGS signal level, and for the case where NGS information
  is only used for tip-tilt correction.}
\label{fig:enc}
\end{figure}

We have selected our primary performance criterion to be the ensquared
energy of the \psf within 75~mas, as motivated by the science
simulations from \citet{Puech2008}, with an updated discussion of
these issues given by \citet{evans2013}.  This spatial scale
(sampled by two \ifu spaxels) was the baseline for the EAGLE concept,
and represents the most demanding requirement on spatial-resolution in
recent studies of a \moao-corrected \mos. As noted in \S\ref{sect:intro},
slightly coarser sampling of 100--150~mas (i.e.\ spatial pixels of 50-75
mas), with comparable ensquared energy, is sufficient to recover the global
properties of high-$z$ galaxies. Such a spaxel size is also a good match
for spectroscopic follow-up of resolved stellar populations observed
with the \hst and, in the future, the \jwst \citep[see discussion
  by][]{evans2013}.

The top-level requirements for MOSAIC demand spaxel sizes which range
from 40 to $\sim$100~mas \citep[see Table 3
  from][]{Evans2014}. Compared to the ensquared energy requirements
from \citet{Puech2008}, the 25--35\% ensquared energy in 75~mas from
the simulated fields in this work satisfy the most demanding
requirements.  Relaxing the spatial scales slightly (to, for example,
the ensquared energy within 150 mas) leads to an ensquared energy
estimate of 40--50\% (Fig.~\ref{fig:enc}). This would provide more
than enough scope for additional reductions in performance from
effects not modelled here (e.g., wavefront errors in the instrument
could potentially degrade the ensquared energy by $\sim$10\%, e.g.,
\citet{2010SPIE.7738E..41L}, while also satisfying the requirements
for the vast majority of the science cases from
\citet{evans2013,Evans2014}.

\subsection{Future work}
We have not considered techniques which make optimum use of \ngs
signals, rather relying on the minimum variance wavefront reconstruction with
\wfs noise covariance approximations.  However, other approaches are
also possible, which we intend to investigate.  This includes the use
of multi-rate \wfss, allowing lower frame rates for \ngs observing
faint targets, and a reduction of \wfs order.  Both techniques would
increase the number of detected photons per sub-aperture per frame,
and would therefore lead to more accurate wavefront slope estimation.

\section{Conclusions}
We have investigated the performance of an \elt \moao instrument,
giving consideration to the effect of \ngs availability on
performance.  We have selected an existing deep cosmological field,
the \goodss field, for this study, which is one of five fields
observed as part of the \hst CANDELS survey, providing rich scientific
potential for \elt spectroscopy.  Ten sub-fields within this region
were chosen at random, and the expected \ao performance investigated.
Within each sub-field, we found at least two sufficiently bright
\ngss, even though a key feature of the deep fields is that they are
deliberately free of bright stars ($V<14$).    We have investigated the \ao
performance as a function of flux from these guide stars, to allow for
uncertainties in detector efficiency and telescope optical throughput.

We find it is beneficial to use high-order wavefront information from
faint natural guide stars as opposed to tip-tilt information. This
conclusion should be revisited once the magnitude of additional
tip-tilt components present within the system (e.g.\ from telescope
vibrations) are better characterised.

Our key \ao performance metric is the ensquared energy within
75~mas, which is between 25-35\% for all of the \ngs asterisms
considered, reducing to about 22\% when no high-order \ngs information
is used.  An increase in ensquared energy to 40-50\% is
possible with a box size of 150~mas.  The \ao performance that we
predict here is sufficient for the proposed MOSAIC instrument.

\section*{Acknowledgements}
This work is funded by the UK Science and Technology Facilities
Council, grant ST/K003569/1.  This research has made use of the APASS
database, located at the AAVSO web site. Funding for APASS has been
provided by the Robert Martin Ayers Sciences Fund.

\bibliographystyle{mn2e}

\bibliography{mybib}
\bsp

\end{document}
